\documentclass{article}
\usepackage{emulateapj}
\usepackage{float}
\usepackage{pstricks}

\begin{document}

\lefthead{}
\righthead{}
\accepted{}
\submitted{Proceedings of the conference {\bf Cosmological 
Constraints from X-ray Clusters}\\
Strasbourg, France, December 9-11, 1998}

\title{Galaxy cluster abundance evolution\\ 
and cosmological parameters}                                               

\medskip

\author{Pedro T. P. Viana}
\affil{Centro de Astrof\'{\i}sica da Universidade do Porto, 
Rua das Estrelas s/n, 4150 Porto, Portugal}
\author{Andrew R. Liddle}
\affil{Astrophysics Group, The Blackett Laboratory,
Imperial College, London SW7 2BZ, United Kingdom}

\begin{abstract}

We use the observed evolution of the galaxy cluster X-ray integral temperature
distribution function between $z=0.05$ and $z=0.32$ in an attempt to
constrain the value of the density parameter, $\Omega_{0}$, for both open and
spatially-flat universes. We conclude that when all the most important sources of 
possible error, both in the observational data and in the theoretical 
modelling, are taken into account, an unambiguous determination of $\Omega_{0}$ 
is not feasible at present. Nevertheless, we find that values of $\Omega_{0}$ 
around 0.75 are most favoured, with $\Omega_{0}<0.3$ excluded with at least 
90 per cent confidence. In particular, the $\Omega_{0}=1$ hypothesis is found 
to be still viable. 

\end{abstract}

\section{Introduction}

The number density of rich clusters of galaxies at the present epoch has been 
used to constrain the amplitude of mass density fluctuations on a 
scale of $8\,h^{-1}\,{\rm Mpc}$ (Evrard 1989; Henry \& Arnaud 1991; 
White, Efstathiou \& Frenk 1993a; Viana \& Liddle 1996; 
Eke, Cole \& Frenk 1996; Kitayama \& Suto 1997). This is usually referred to 
as $\sigma_{8}$, where $h$ is the present value of the Hubble parameter, 
$H_{0}$, in units of $100\;{\rm km}\,{\rm s}^{-1}\,{\rm Mpc}^{-1}$. 
However, the derived value of $\sigma_{8}$ depends to a great extent on 
the present matter density in the Universe, parameterized by $\Omega_{0}$, 
and more weakly on the presence of a cosmological constant, $\Lambda$. 
The cleanest way of breaking this degeneracy is to include information 
on the change in the number density of rich galaxy clusters with 
redshift (Frenk et al. 1990), the use of X-ray clusters for this purpose 
having been proposed by Oukbir \& Blanchard (1992) and subsequently further 
investigated (Hattori \& Matsuzawa 1995; Oukbir \& Blanchard 1997). 
Several attempts have been made recently, with wildly differing results 
(Henry 1997; Fan, Bahcall \& Cen 1997; 
Blanchard \& Bartlett 1997; Eke et al. 1998; Gross et al. 1998; 
Reichart et al. 1998; Viana \& Liddle 1999).

The best method to find clusters of galaxies is through their X-ray emission,
which is much less prone to projection effects than optical identification.
Further, the X-ray temperature of a galaxy cluster is at present the most
reliable estimator of its virial mass.  This can then be used to relate the
cluster mass function at different redshifts, calculated for example within
the Press--Schechter framework (Press \& Schechter 1974; Bond et al.  1991), 
to the observed cluster X-ray
temperature function. We can therefore compare the evolution in the number
density of galaxy clusters seen in the data with the theoretical expectation
for large-scale structure formation models, which depends significantly only 
on the assumed values of $\Omega_{0}$ and $\lambda_{0}\equiv\Lambda/3H^{2}_{0}$, 
the latter being the contribution of $\Lambda$ to the total present energy 
density in the Universe.

We used two complete X-ray samples of galaxy clusters to estimate the 
evolution in their number density between the redshifts of $z=0.05$ and 
$z=0.32$. The low-redshift X-ray sample is an updated version, kindly provided 
to us by Pat Henry, of the dataset first presented in Henry \& Arnaud (1991). 
We used it to to estimate $\sigma_{8}$ as a function of $\Omega_{0}$ for the 
cases of of an open universe, where the cosmological constant is zero, 
and a spatially-flat universe, such that $\lambda_{0}=1-\Omega_{0}$. All 
the dark matter was assumed cold. The calculation was performed within the 
extended Press--Schechter formalism proposed by Lacey \& Cole (1993, 1994), 
which allows an estimation of the formation times of the dark matter halos 
associated with the galaxy clusters, and updates the result of 
Viana \& Liddle (1996). Knowing $\sigma_{8}$ for each $\Omega_{0}$ model 
allows the estimation of the expected number density of galaxy clusters  
model at any redshift as a function of $\Omega_{0}$. Hence, comparing with 
high-redshift data one is then 
able to find the $\Omega_{0}$ model which predicts cluster numbers closest 
to the observed ones. 

The high-redshift X-ray sample we considered is made of the 10 
clusters with redshifts between 0.3 and 0.4\footnote{Recently, it has been 
found that one of these clusters, MS1241.5, actually has a redshift of 
0.549. Our results do not change significantly.} (the median redshift 
being 0.32) and with X-ray fluxes in excess of 
$2.5\times10^{-13}\;{\rm erg}\,{\rm cm}^{-2}\,{\rm s}^{-1}$, that were 
identified in the {\em Einstein Medium Sensitivity Survey} ({\em EMSS}) 
(Gioia et al. 1990; Henry et al. 1992; Gioia \& Luppino 1994; 
Nichol et al. 1997). For the X-ray fluxes and temperatures of these 
clusters we used the values published in Henry (1997), which were 
based on {\em ASCA} data. 

We chose to concentrate on clusters with X-ray temperatures in excess 
of 6.2 keV. The value of the integral X-ray temperature function at this 
point is closest to the mean curves that best represent the integral X-ray 
temperature functions both at $z=0.05$ and $z=0.32$, recovered from the 
observational data. And considering only the most massive clusters one 
has several advantages: the underlying density field on the 
scales associated with these clusters still retains to a great extent its 
assumed initial Gaussianity, which is the main requirement 
to apply the Press--Schechter ansatz; as these clusters represent the 
highest peaks in the density field their number 
density is very sensitive to the normalization of the perturbation 
spectrum, which in turn means that their number density is expected to 
change rapidly as we move into the past; the X-ray temperature of these 
clusters has probably been little affected by a possible pre-heating of the 
intracluster medium (e.g. Metzler \& Evrard 1994; Navarro, Frenk \& White 1995; 
Cavaliere, Menci \& Tozzi 1997)

\section{Models versus data}

In Viana \& Liddle (1999) we describe in detail the steps we took to 
estimate both the theoretically-expected value for the number density of 
galaxy clusters with X-ray temperatures in excess of 6.2 keV at $z=0.32$, 
and its observational counterpart based on the data in Henry (1997). 

\subsection{What models predict}

The theoretical estimation was performed for different values of 
$\Omega_{0}$, and in the two cases of open and spatially-flat models. 
The input observational data used in this calculation was the the number density 
of galaxy clusters with X-ray temperatures in excess of 6.2 keV at $z=0.05$, 
obtained through the updated version of the dataset in Henry \& Arnaud (1991), 
and the shape of the galaxy and cluster power spectrum, which was assumed to be 
well described in the vicinity of $8\,h^{-1}\,{\rm Mpc}$ by a scale-invariant 
cold dark matter power spectrum with a shape parameter of 
$\Gamma=0.230^{+0.021}_{-0.017}$ where the quoted error is $1\sigma$ 
(Peacock \& Dodds 94; Viana \& Liddle 96). In short, we used the 
Press--Schechter formalism to calculate the cluster mass function, which was 
then transformed in a cluster X-ray temperature function via a virial mass 
to X-ray temperature relation normalized through data from hydrodynamical 
N-body simulations (White et al. 1993b; Metzler \& Evrard 1994; 
Navarro, Frenk \& White 1995; Bryan \& Norman 1998), 
and where the formation redshift of the clusters was 
estimated by means of the extension to the Press--Schechter framework 
proposed by Lacey \& Cole (1993, 1994). Besides the errors in the observed 
values for $N(>6.2\,{\rm keV},\,z=0.05)$ and $\Gamma$, 
in this calculation there were three more sources of uncertainty: 
the normalization of the virial mass to X-ray temperature relation has 
a $1\sigma$ associated error of around 25 per cent in the 
mass at fixed temperature; the threshold value 
that characterizes virialized structures in the Press-Schechter ansatz, 
was assumed to equal $\delta_{c}=1.7\pm0.1$ at the $1\sigma$ level of confidence; 
and the fraction of the mass of a cluster that needs to be assembled before it can 
be considered to have formed (in the sense that its X-ray properties do not change 
much henceforth), was taken to be $f=0.75\pm0.075$, also at the $1\sigma$ level. 

The input observed value we used for $N(>6.2\,{\rm keV},\,z=0.05)$ 
was previously corrected from the bias introduced due to 
X-ray temperature measurement errors. In the usual situation 
where the instrumental errors are to a first approximation symmetrical, 
it is as probable for a cluster with actual X-ray 
temperature below 6.2 keV to have a measured temperature above that value, 
as it is for a cluster with $k_{\rm B}T>6.2$ keV
to have a smaller measured temperature. However given that the cluster number 
density increases as the cluster X-ray temperature decreases, there will be a net 
effect, due to the presence of the temperature measurement errors, 
leading to an apparent increase in the number density of galaxy clusters 
with X-ray temperature above 6.2 keV. This overestimation of the observed value 
for $N(>6.2\,{\rm keV},\,z=0.05)$ was corrected by using the mean for this quantity 
calculated from 10000 bootstrap samples, assembled by randomly selecting, 
with replacement, from the original list of 25 X-ray clusters in 
Henry \& Arnaud (1991), and where each time a cluster
was selected its X-ray temperature was randomly drawn from a
Gaussian distribution with the mean and dispersion observed for the cluster. 
The mean obtained in this way is larger than the value 
for $N(>6.2\,{\rm keV},\,z=0.05)$ extracted from the original dataset, in the same 
proportion as this value is larger than the best estimate for the actual 
$N(>6.2\,{\rm keV},\,z=0.05)$ in the Universe. In all, the corrected best estimate 
for the number density of galaxy clusters at $z=0.05$ with X-ray temperature 
exceeding 6.2 keV is
\begin{equation}
N(>6.2\,{\rm keV},\,0.05)=1.53\times10^{\pm0.16}\times10^{-7}h^{3}\;{\rm Mpc}^{-3}
\end{equation}
where the errors represent 1-sigma confidence levels. These were 
obtained through the bootstrap procedure, which allows an estimation of the 
uncertainty associated with the sampling variance. The uncertainty 
associated with the counting error (i.e. cosmic variance) was also included 
by drawing the number of clusters in each bootstrap sample from a Poisson 
distribution with mean 25. 

As a by-product of the calculation of the theoretically-expected value for 
$N(>6.2\,{\rm keV},\,z=0.32)$ we obtained a new estimate for $\sigma_{8}$, 
as a function of both $\Omega_{0}$ and $\lambda_{0}$, so that the 
corrected observed value for $N(>6.2\,{\rm keV},0.05)$ is reproduced. 
This supersedes the result obtained in Viana \& Liddle (1996). 
We find that the best-fitting value is given by
\begin{displaymath}
\label{final1}
\sigma_8=\left\{ \begin{array}{ll}
0.56 \; \Omega_0^{-0.34} & {\rm Open}\\
0.56 \; \Omega_0^{-0.47} & {\rm Flat}
\end{array}\right.
\end{displaymath}
with an accuracy better than 3 per cent for $\Omega_{0}$ between 0.1 and 1. 
 
The most important reason why this value is smaller than that quoted in 
Viana \& Liddle (1996) 
is the decrease in the assumed number density of galaxy clusters at $z=0.05$. 
This results from the revision of the Henry \& Arnaud dataset and from 
the correction due to the existence of X-ray temperature measurement 
errors, which had not been taken into consideration in 
Viana \& Liddle (1996) . Also, the 
cluster X-ray temperature function obtained in Henry \& Arnaud (1991)
had been slightly overestimated due to a calculational error (Eke, Cole and Frenk 1996). 

The overall uncertainty in the value of $\sigma_{8}$ and therefore also in the 
theoretically expected value for 
$N(>6.2\,{\rm keV},\,z=0.32)$ was calculated in the 
same way as in Viana \& Liddle (1996), through a Monte Carlo procedure where the 
sources of error, namely the normalization of the cluster X-ray temperature to virial 
mass relation, the value of $\delta_{{\rm c}}$ and the value of $f$, are 
modeled as being Gaussian distributed, and the observed $\Gamma$ and 
$N(>6.2\,{\rm keV}, 0.05)$ as having a lognormal distribution. As in 
Viana \& Liddle (1996) 
we find that for each $\Omega_{0}$ between 0.1 and 1 the distribution of 
$\sigma_{8}$ can be approximated by a lognormal. For open models, the 
95 per cent confidence limits are roughly given by $+20\Omega_0^{0.1{\rm 
log}_{10} \Omega_0}$ per cent and $-18\Omega_0^{0.1{\rm log}_{10} 
\Omega_0}$ per cent, while for flat models we have $+20\Omega_0^{0.2{\rm 
log}_{10} \Omega_0}$ per cent and $-18\Omega_0^{0.2{\rm log}_{10} \Omega_0}$. 

\subsection{What the observations tell us}

The observed value for the number density of galaxy clusters with X-ray 
temperatures in excess of 6.2 keV at $z=0.32$ was calculated by using the 
data in Henry (1997). The estimator used was
\begin{equation}
\label{est}
N(>k_{{\rm B}}T,\,z)=\sum_{i=1}\frac{1}{V_{{\rm max},i}}
\end{equation}
where the sum is over all clusters with 
$k_{{\rm B}} T_{i}>k_{{\rm B}}T$, and $V_{{\rm max},i}$ is the maximum volume 
in which cluster $i$ could have been detected at the $4\sigma$ level in the 
{\em EMSS} within the redshift shell under consideration (0.3 to 0.4 in our
case). The steps which need to be taken in order to calculate these volumes are
described in Henry et al. (1992) and Viana \& Liddle (1999). 

As in the lower redshift case, the calculation of 
$N(>6.2\,{\rm keV},\,0.32)$ using the X-ray temperatures measured for 
the galaxy clusters found between redshifts 0.3 and 0.4 would lead to the
overestimation of the true $N(>6.2\,{\rm keV},\,0.32)$ due to the presence of
errors in the X-ray temperature determinations.  Again, we corrected for
this by simulating the repetition of the X-ray temperature measurements 
through a bootstrap procedure analogous to the one performed for the 
low-redshift data, though in this case the number of clusters in each sample 
was now extracted from a Poisson probability distribution with mean 10. When 
a cluster is selected for a simulated dataset we not only let its X-ray 
temperature vary within the observational errors, but also its flux, which 
affects the maximum search volume associated with the cluster. 
The ratio between the mean value obtained for 
$N(>6.2\,{\rm keV},\,0.32)$ from all the simulated datasets, and the value 
one gets for $N(>6.2\,{\rm keV},\,0.32)$ using the original dataset, then 
provides an estimate for the expected ratio between the latter and the 
real value for $N(>6.2\,{\rm keV},\,0.32)$ in the Universe. The 
uncertainty in this value is obtained from the results of the bootstrap 
procedure, which in practice simulates the repetition 
a large number of times across the sky of the type of sampling 
that led to the dataset in Henry (1997). 

\subsection{The comparison}

Let us now assume that in our Universe $N(>6.2\,{\rm keV},\,0.32)$
takes some particular overall value, the one theoretically-expected given 
the $\Omega_{0}$ model under consideration. We would then expect this value
to be the {\em mean} of the distribution function assembled with the
values that would be measured for $N(>6.2\,{\rm keV},\,0.32)$ if the
type of sampling that led to the dataset in Henry (1997) 
was repeated a large number of times across the sky.  On the other
hand, we would expect that the {\em shape} of this distribution would
be that obtained through the bootstrap procedure mentioned at the end
of the previous subsection.  We are therefore now in a position to ask
the following question. If $N(>6.2\,{\rm keV},\,0.32)$ took such an 
overall value in the Universe, how probable would it be for an 
observer to measure the
value for $N(>6.2\,{\rm keV},\,0.32)$ given by the dataset in Henry
(1997), after correcting it for the displacement due to
errors in the X-ray temperature measurements. We can then attach, for
each value of $\Omega_{0}$, a probability of the value for
$N(>6.2\,{\rm keV},\,0.32)$ given by the dataset in Henry (1997) 
being actually measured. The {\em exclusion level}
on each value of $\Omega_{0}$ is obtained simply by subtracting
this probability from one.

In summary the following steps were taken, so that an exclusion level
can be associated with each $\Omega_0$ based on the X-ray cluster
datasets for $z=0.05$ and $z=0.32$:
\begin{enumerate}
\item The Universe was assumed to be either open or spatially-flat, 
with $\Omega_{0}$ taking a value between 0.1 and 1.
\item The best estimate for $N(>6.2\,{\rm keV},\,0.32)$ in the Universe, 
given the dataset in Henry (1997), was calculated taking into 
consideration the effect of the X-ray temperature measurement errors. 
\item A bootstrap procedure analogous to that described for 
the $z=0.05$ data was performed in order to determine the expected 
{\em shape} for the distribution function of $N(>6.2\,{\rm keV},\,0.32)$,
if the type of sampling that led to the dataset in Henry (1997) 
was repeated a large number of times across the sky. The number of clusters 
in each sample is now drawn from a Poisson distribution with mean 10, and  
the input observational errors (most importantly the {\em ASCA} X-ray fluxes 
and temperatures) are modeled as Gaussian distributed. 
\item Through the extended Press-Schechter formalism, the 
theoretically-expected overall value for $N(>6.2\,{\rm keV},\,0.32)$ 
given the assumed $\Omega_{0}$ was calculated. The normalization
$\sigma_8$ of the spectrum was fixed by the low-redshift data.
\item The distribution function for $N(>6.2\,{\rm keV},\,0.32)$, determined 
through the bootstrap procedure, was modified by dividing the values obtained 
for $N(>6.2\,{\rm keV},\,0.32)$ by their mean and multiplying them by the 
value determined in (iv), so that this value becomes the new mean and the 
relative shape of the distribution is maintained. 
\item We calculated the probability of obtaining a value as high, or as low, 
as that determined in (ii), given the distribution constructed in (v). The 
exclusion level on the assumed $\Omega_{0}$ equals one minus this 
probability.  
\end{enumerate}

However, due to the uncertainties in the estimation of the
theoretically-expected overall value for $N(>6.2\,{\rm keV},\,0.32)$,
the actual calculation of the exclusion level for each $\Omega_{0}$ was 
in reality slightly more complicated. So that we could obtain it, we needed to 
integrate over all possible values for the theoretically-expected 
$N(>6.2\,{\rm keV},\,0.32)$, which we will denote $u$. The overall exclusion 
is the product of the probability, $P(u,\Omega_{0})$, of each $u$ being
the correct overall value one would expect for 
$N(>6.2\,{\rm keV},\,0.32)$ in the Universe (given the assumed $\Omega_{0}$), 
and the exclusion level ${\rm Ex}(u)$ calculated as described above for each 
assumed $u$, i.e.
\begin{equation}
\label{excl}
\hspace{-0.2cm}{\rm Exclusion}\,{\rm probability}\,{\rm of}\,\Omega_{0}=
\int_{-\infty}^{+\infty}P(u,\Omega_{0})\,{\rm Ex}(u)du
\end{equation}
The $P(u,\Omega_{0})$ are lognormal distributions with 
mean equal to the value calculated in (iv), and dispersion obtained 
through Monte Carlo simulations mentioned at the end of subsection (2.1).

In Figure~1 we show the exclusion levels for $\Omega_{0}$ obtained in
this way.  Even for the values of $\Omega_{0}$ for which it is easiest
to reproduce the observations, from 0.7 to about 0.8, the exclusion
level is quite high, around 70 per cent.  The reason lies with the
large uncertainty in the theoretically-expected overall value for
$N(>6.2\,{\rm keV},\,0.32)$.  Because of it, most
theoretically-expected values end up far away from the value
for $N(>6.2\,{\rm keV},\,0.32)$ obtained from 
the dataset in Henry (1997). A large uncertainty in
the theoretical prediction is clearly no basis to discard
models. However, for the high and low $\Omega_{0}$ we are aiming to
constrain, this effect becomes much less important; the high exclusion
levels are caused by most of the distribution for the
theoretically-expected overall values for $N(>6.2\,{\rm keV},\,0.32)$
being higher (for low $\Omega_0$), or lower (for high $\Omega_0$),
than the observations. Note that the exclusion levels are absolute,
and not relative as one would obtain from the calculation of a
likelihood function.

\bigskip
\centerline{\includegraphics[width=3.25in]{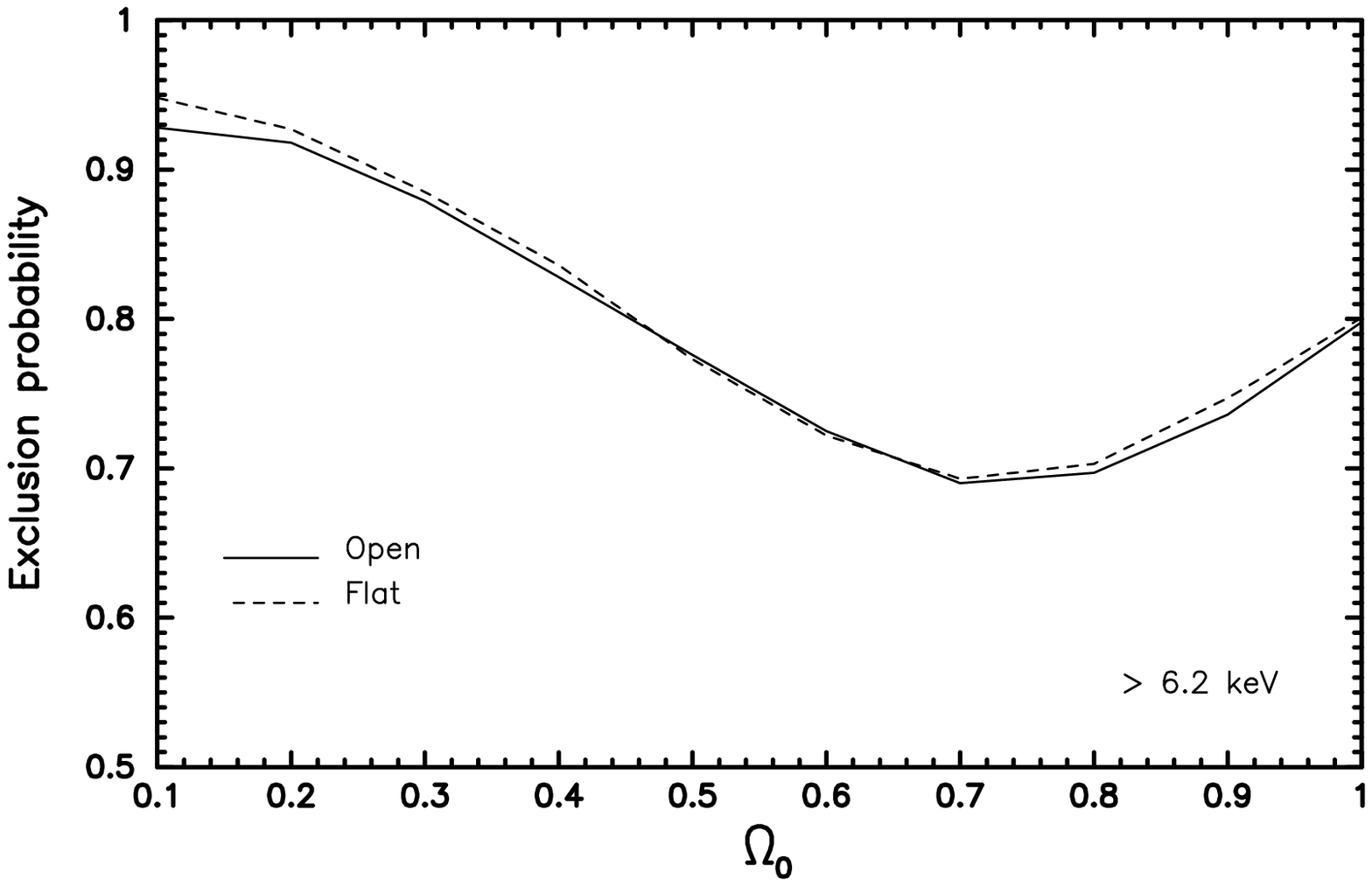}}
\figcaption{The absolute exclusion levels for 
different values of $\Omega_{0}$ in both the open and 
spatially-flat cases, when a threshold X-ray temperature of
6.2 keV is used.}  
\medskip

The calculations we have just described assume that there is no scatter in
the relation between cluster X-ray temperature and luminosity.  This is not
correct and can lead to an {\em increase} in the
observed value for $N(>k_{\rm B}T,\,z)$.  This incompleteness problem
worsens as the threshold X-ray temperature $k_{\rm B}T$ is lowered, as one 
then starts considering clusters with X-ray fluxes dangerously close to the flux
detection limit.  For the same threshold X-ray temperature, the problem is
also potentially much more serious in the case of the $z=0.32$ data than for
the $z=0.05$ data.  The reason is simply that for the same flux detection
limit, the faintest clusters that can be detected nearby have X-ray
luminosities (and thus temperatures) which are considerably smaller than
those of the faintest clusters further away.

While the effect of the scatter in the X-ray cluster temperature--luminosity
relation in the calculation of $N(>6.2\,{\rm keV},\,0.05)$ is negligible, as
the $z=0.05$ dataset in Henry \& Arnaud (1991) is claimed to be
nearly complete down to at least 3 keV, in the case of the $z=0.32$ data 
it is not clear whether the presence of scatter in the X-ray cluster 
temperature--luminosity relation may affect the determination of 
$N(>6.2\,{\rm keV},\,0.32)$. Due to the scatter,
there is a finite probability that some of the 5 {\em EMSS}
galaxy clusters with X-ray flux below $2.5\times10^{-13}\;{\rm erg}\,{\rm
cm}^{-2}\,{\rm s}^{-1}$, that were found in the redshift range from 0.3 to 
0.4 (Henry et al. 1992), may not only have an X-ray temperature in excess of 
the lowest X-ray temperature present in the sub-sample of 10 clusters from 
Henry (1997), 3.8 keV for MS1512.4, but also in excess of our chosen 
threshold temperature 6.2 keV.  The minimization of this possibility 
was in fact another reason for our choice of 6.2 keV as the threshold 
temperature. 

We calculated the expected increase in the corrected value of
$N(>6.2\,{\rm keV},\,0.32)$ given the dataset in Henry (1997), as
a result of the existence of the 5 {\em EMSS} galaxy clusters mentioned
above, by doing 1000 Monte Carlo simulations where the X-ray temperatures for
those clusters were estimated via the X-ray cluster temperature--luminosity
relation determined in Eke et al. (1998) using the more recent 
data for the galaxy clusters in Henry \& Arnaud (1991). Though this relation 
is that observed for galaxy clusters at $z=0.05$, the recent analyses of 
Mushotzky \& Scharf (1997) and Allen \& Fabian (1998) (see
also Sadat, Blanchard and Oukbir 1998) seem to imply it holds at least 
up to $z=0.4$.  

In the end, we found that allowing for the presence of scatter in the X-ray
cluster temperature--luminosity relation when calculating $N(>6.2\,{\rm
keV},\,0.32)$ has only a small effect, at the few per cent level, 
on the exclusion levels obtained for different $\Omega_{0}$, and does 
not alter our conclusions. 

We would also like to draw attention to the fact that 
in all previous uses of the Press--Schechter framework to calculate
the evolution of the number density of rich galaxy clusters with
redshift 
(Oukbir \& Blanchard 1992; Eke, Cole \& Frenk 1996; Oukbir \& Blanchard 1997; 
Henry 1997; Eke et al. 1998; Markevitch 1998; Reichart et al. 1998), 
it has been assumed that the redshift of cluster virialization, $z_{{\rm c}}$,
coincides with that at which the galaxy cluster is observed, $z_{\rm
obs}$.  In Figure 2 we compare the value of $N(>6.2\,{\rm keV},\,z)$
obtained using the Lacey \& Cole (1993, 1994) prescription for the
estimation of $z_{{\rm c}}$ with the result of the assumption that $z_{{\rm
c}}=z_{{\rm obs}}$.  We always require that the observed value for
$N(>6.2\,{\rm keV},\,0.05)$ is recovered.

\bigskip
\centerline{\includegraphics[width=3.25in]{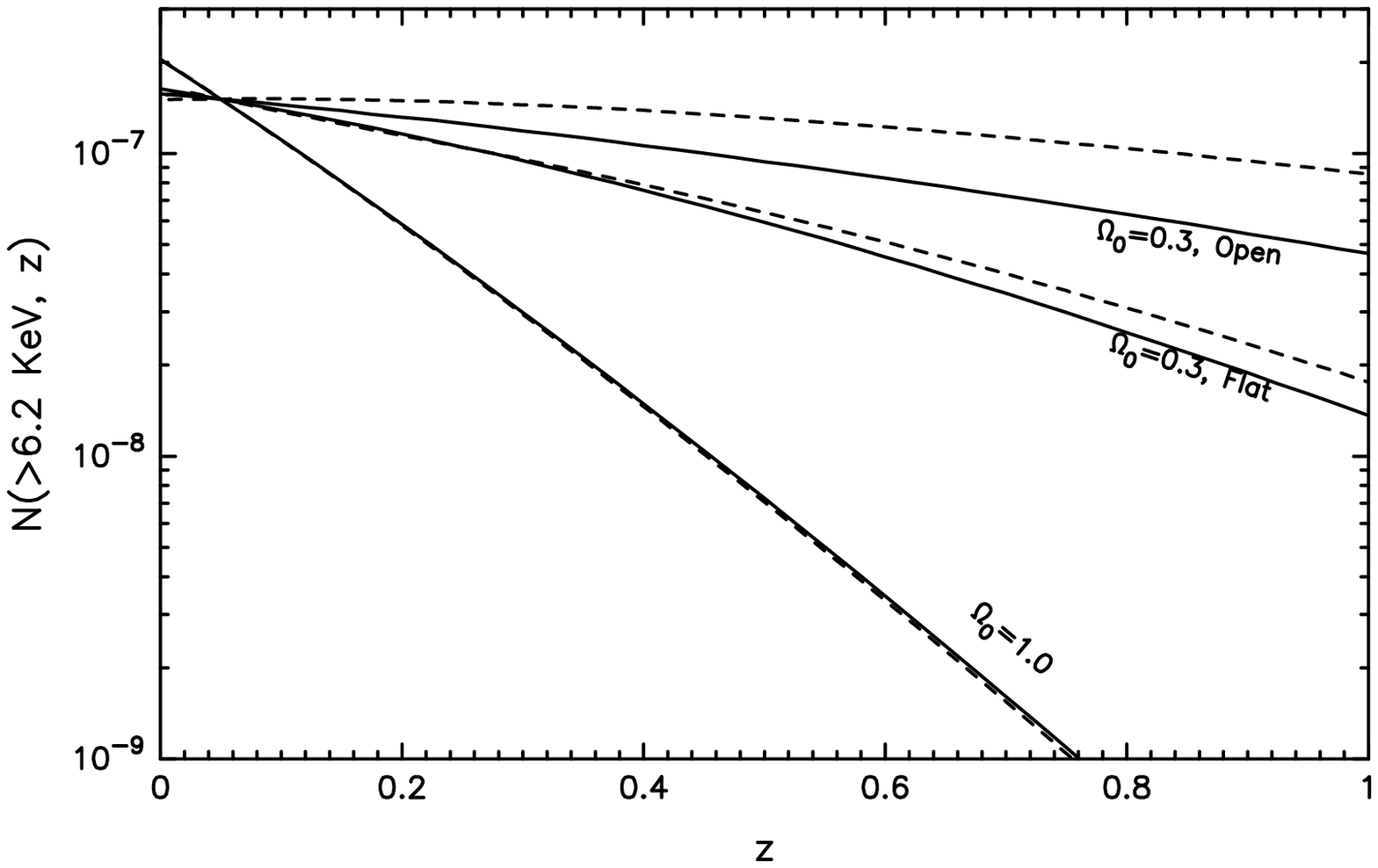}}
\figcaption{The expected redshift evolution of 
$N(>6.2\,{\rm keV},\,z)$ for $\Omega_{0}=1$ and 0.3. The solid lines show the 
result obtained using the Lacey \& Cole method for estimating $z_{{\rm c}}$, 
and the dashed ones the result obtained assuming that $z_{{\rm c}}=z_{{\rm 
obs}}$. Each curve is normalized to reproduce the observed value for 
$N(>6.2\,{\rm keV},\,0.05)$. Note that the divergence at high $z$ is caused 
by this renormalization; the absolute correction is largest at the lowest 
redshift, where $\Omega(z)$ is smallest.}  
\medskip

As expected, the difference in the theoretically-predicted overall 
value of $N(>6.2\,{\rm keV},z)$ 
resulting from the two distinct assumptions regarding $z_{{\rm c}}$ becomes
larger for $\Omega_{0}=0.3$, reflecting the fact that as $\Omega_{0}$ goes
down galaxy clusters tend to form increasingly at an earlier epoch than that
at which they are observed.  We found that neglecting the fact that some
clusters of galaxies virialize prior to the epoch at which they are observed
leads to an underestimation of the predicted degree of evolution
in the value of $N(>k_{\rm B}T,\,z)$ for $z>z_{\rm norm}$, where $z_{\rm
norm}$ is the redshift at which $N(>k_{\rm B}T,z)$ is normalized through
observations, e.g in our case $z_{\rm norm}=0.05$.  Taking into account the
possibility that $z_{{\rm c}}$ may be larger than $z_{\rm obs}$ therefore
requires {\em lower} values for $\Omega_{0}$ in order for the high-redshift
data on $N(>k_{\rm B}T,\,z)$ to be reproduced.

Allowing for $z_{{\rm c}}>z_{{\rm obs}}$ means that some galaxy clusters that
otherwise would not be massive enough to reach a given threshold temperature
$k_{\rm B}T$ can now be counted when calculating $N(>k_{\rm B}T,\,z)$.  In
principle this would have the effect of increasing the expected value of
$N(>k_{\rm B}T,\,z)$ for any $z$.  However, at the normalization redshift
0.05 the higher value for $N(>k_{\rm B}T,\,0.05)$ means that a less well
developed density field at $z=0.05$ is required, i.e.~a lower value of
$\sigma_{8}$ results from introducing the possibility that $z_{{\rm
c}}>z_{\rm obs}$.  As the number density of virialized objects evolves faster
for the same relative change in the value of the dispersion of the density
field the smaller this value is, the decrease in the required value for
$\sigma_{8}$ has the effect of enhancing the decrease in the value of
$N(>k_{\rm B}T,\,z)$ as $z$ gets larger.  This effect turns out to be more
important than the expected increase in the value of $N(>k_{\rm B}T,\,z)$ due
to higher cluster X-ray temperatures at fixed cluster mass resulting from the
possibility of $z_{{\rm c}}>z_{\rm obs}$.

\section{Discussion}

From the above analysis, we conclude that {\em at present} it is not
possible to reliably exclude any interesting value for $\Omega_{0}$ on
the basis of X-ray cluster number density evolution alone, due to the
limited statistical significance of the available observational data
and to uncertainties in the theoretical modelling of cluster formation
and evolution.  However, we do find that values of $\Omega_{0}$ below
0.3 are excluded at least at the 90 per cent confidence level.  Values
of $\Omega_{0}$ between 0.7 to 0.8 are those most favoured, though not
strongly.  These results are basically independent of the presence or
not of a cosmological constant.

Our conclusions support those of Colafrancesco, Mazzotta \& Vittorio (1997), 
who tried to estimate the uncertainty involved in the 
estimation of the cluster X-ray temperature distribution function at 
different redshifts based on its present-day value. They found this 
uncertainty, given the still relatively poor quality of the data, to be 
sufficiently large to preclude the imposition of reliable limits on the 
value of $\Omega_{0}$. 
 
Our results disagree with those of Henry (1997) and 
Eke et al. (1998), as they found the preferred $\Omega_{0}$ 
to lie between 0.4 to 0.5, with the $\Omega_{0}=1$ hypothesis strongly 
excluded. This disagreement is mainly the consequence of our focus on the 
threshold X-ray temperature of 6.2 keV, while they draw their conclusions 
based on the analysis of the results obtained for several threshold X-ray 
temperatures. Other less important contributions to the difference 
between our results and theirs are the different assumed 
normalization for the virial mass to X-ray temperature relation, 
and the corrections in the expected values in the Universe for 
both $N(>6.2\,{\rm keV},\,0.05)$ and $N(>6.2\,{\rm keV},\,0.32)$ due to 
the uncertainties in the X-ray cluster temperature measurements. 
Our disagreement with Eke et al. (1998) on the level of 
exclusion of the $\Omega_{0}=1$ hypothesis is also due to our much larger 
assumed uncertainty in the theoretically-expected overall value for 
$N(>6.2\,{\rm keV},\,0.32)$. 

In Figure 3 we show the absolute exclusion levels for different values 
of $\Omega_{0}$ in both the open and spatially-flat cases, when a threshold 
X-ray temperature of 4.8 keV is used. They are substantially different from 
those we obtained when the threshold X-ray temperature was assumed to be 6.2 keV.  
We found this to be even more true if a correction for the possibility of any of 
the 5 clusters with the lowest X-ray fluxes in the $0.3<z<0.4$ EMSS sub-sample 
having X-ray temperatures in excess of 4.8 keV is included. While the standard 
analysis without these 5 X-ray clusters prefers a value
for $\Omega_{0}$ between 0.4 to 0.5, when the correction for the scatter in
the relation between the cluster X-ray temperature and luminosity is
included, in the way described in subsection 2.3, the preferred value for
$\Omega_{0}$ decreases to about 0.3.  Now the $\Omega_{0}=1$ hypothesis is
excluded at more than the 95 per cent confidence level, with or without the
correction.  At the 90 per cent confidence level, one finds that
$\Omega_{0}>0.8$ is excluded without the correction, being this limit lowered
to 0.7 when the correction is included. 

\bigskip
\centerline{\includegraphics[width=3.25in]{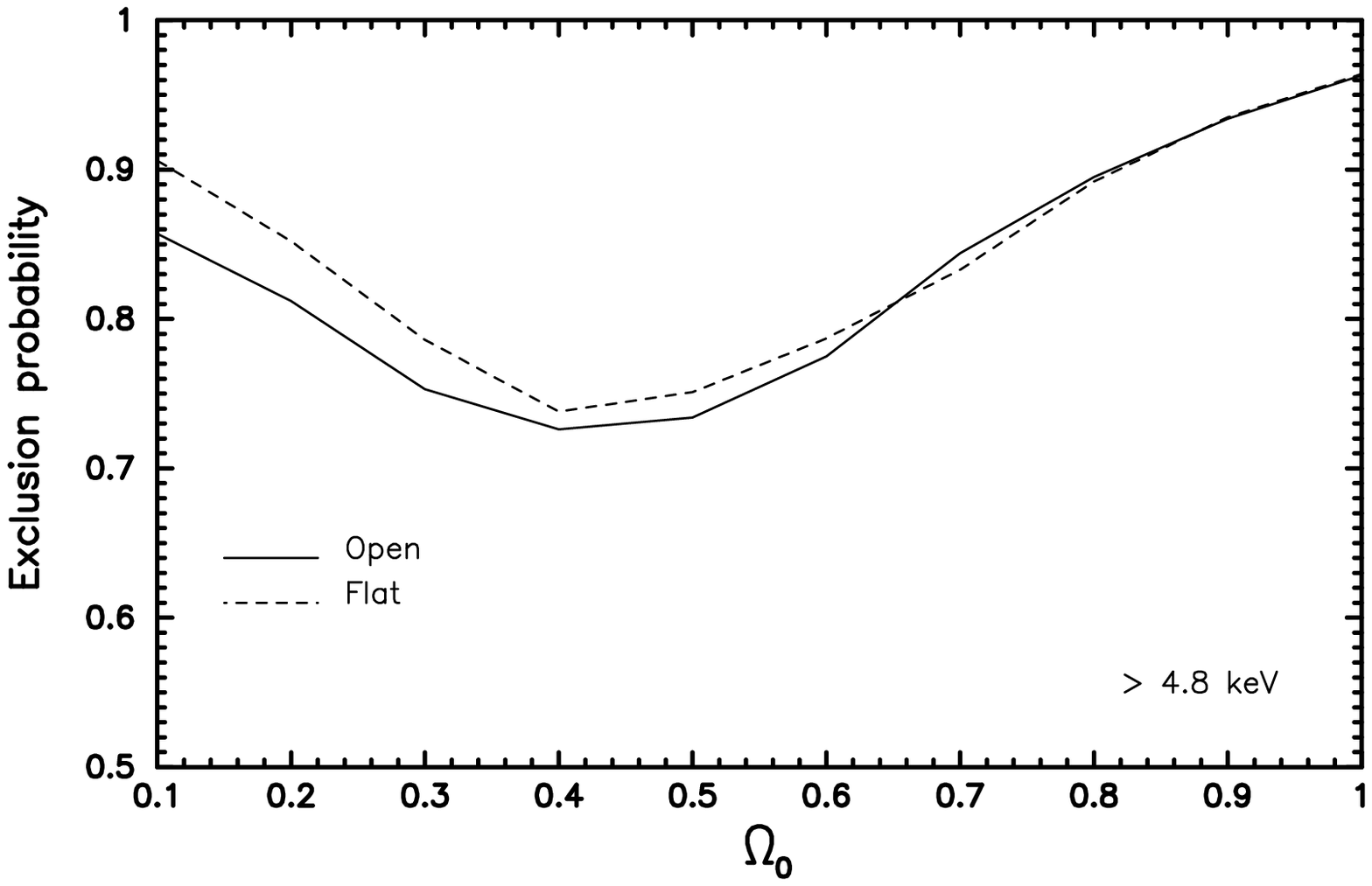}}
\figcaption{The absolute exclusion levels for 
different values of $\Omega_{0}$ in both the open and 
spatially-flat cases, when a threshold X-ray temperature of
4.8 keV is used.}  
\medskip

One can also estimate the joint probability of some $\Omega_{0}$ value being 
excluded on the basis of the results relative to either one or both  
X-ray temperature thresholds. Assuming the data used in the 
calculations for the two thresholds is independent, the results then 
imply that the favoured value for $\Omega_{0}$ is close to 0.55 
(0.50 if the incompleteness correction is included) and the 
$\Omega_{0}=1$ hypothesis is excluded at the 99 per cent level. This agrees 
very well with the results of Henry (1997) and 
Eke et al. (1998), leading us to believe that the main 
difference between our analysis and theirs was our decision to 
draw our conclusions solely based on the exclusion levels obtained for the 
X-ray temperature threshold of 6.2 keV. Nonetheless, we feel that only the 
data regarding clusters with X-ray temperatures in excess of about 6 keV seems
sufficiently free of modelling problems, like sample incompleteness and 
significant pre-heating of the intracluster medium, so as to be potentially 
useful in constraining $\Omega_{0}$, at least as long as these issues are 
not satisfactoroly settled. An example of possible sample incompleteness 
has arisen from recent work by Blanchard, Bartlett and Sadat (1998), who 
used a sample of 50 galaxy clusters with mean redshift of 0.05, which were 
identified through the {\em ROSAT} satellite, to estimate the cumulative X-ray 
temperature distribution function at $z=0.05$. They claim the number density 
of galaxy clusters at $z=0.05$ with X-ray temperatures around 4 keV 
is being {\em underestimated} when the Henry \& Arnaud cluster sample is 
used. Through the X-ray cumulative temperature distribution function 
at $z=0.05$ they obtain, they then estimate $\Omega_{0}$ using the 
{\em EMSS} cluster abundance in the redshift bin $0.3<z<0.4$ and the X-ray 
temperature data gathered in Henry (1997). They find 
the favoured value for $\Omega_{0}$ to be 0.75, while $\Omega_{0}<0.3$ is 
excluded at more than the 95 per cent level. These results coincide 
very well with ours, when only the 6.2 keV threshold X-ray temperature 
is considered, thus perhaps implying that the discrepancy between the 
favoured value for $\Omega_{0}$ found when different X-ray temperature thresholds 
are considered may arise from a underestimation of the cumulative distribution 
function at $z=0.05$ for X-ray temperatures below about 6 keV. 

So at the moment the situation is that, 
unfortunately, due to uncertainties associated both with the observational 
measurements and the theoretical modelling of cluster evolution, 
the presently-available X-ray data on galaxy clusters is not able to strongly
discriminate between cosmologies with different values for $\Omega_{0}$.  
And in any case, the data available is probably not yet statistically
significant. More is needed to support or disclaim the preliminary
conclusions that can be obtained from it. 

\section{Future prospects}

Within the next few years, with the launch of the {\em XMM} satellite,  
possibly in early 2000, a significant increase in the quantity and quality 
of the available data is expected to occur (Romer 1998). It should then be 
possible to place stronger constraints on $\Omega_{0}$ 
on the basis of the evolution of the galaxy cluster X-ray temperature 
function. This would be helped by improvements in the theoretical 
modelling of cluster evolution, perhaps based on the high-resolution 
hydrodynamical $N$-body simulations on cosmological scales expected 
in the near future.

In Figure 4 we show predictions for the 
cumulative redshift distribution, both out to and beyond some redshift $z$, of 
the number of galaxy clusters per 1000 square degrees in the sky 
with X-ray temperature in excess of 6 keV. The results were obtained for 
two models with $\Omega_{0}=0.3$, one open and the other spatially-flat, and 
a model with $\Omega_{0}=1$. The solid lines include all 
clusters with such temperatures, independently of their X-ray flux.  
The dashed and dotted lines represent only the clusters which have a X-ray flux 
in the [2,10] keV band in excess of 
$10^{-13}$ and $10^{-12}\;{\rm erg}\,{\rm cm}^{-2}\,{\rm s}^{-1}$, 
respectively. The conversion of X-ray temperature to flux was performed using 
the X-ray temperature to luminosity relation derived by Allen \& Fabian (1998) 
for high-temperature clusters with redshifts up to 0.4. If this relation breaks 
down for higher redshifts, the cluster abundance numbers presented will change 
beyond roughly the redshift at which the predicted X-ray flux in the [2,10] keV 
band for a 6 keV cluster equals the threshold fluxes quoted. This means that the 
$10^{-12}\;{\rm erg}\,{\rm cm}^{-2}\,{\rm s}^{-1}$ results would be little 
affected given that very few clusters above $z=0.4$ are predicted to have such 
a high X-ray flux, while the $10^{-13}\;{\rm erg}\,{\rm cm}^{-2}\,{\rm s}^{-1}$ 
results would only be affected above a redshift of about 1. 

\bigskip
\centerline{\includegraphics[width=3.25in]{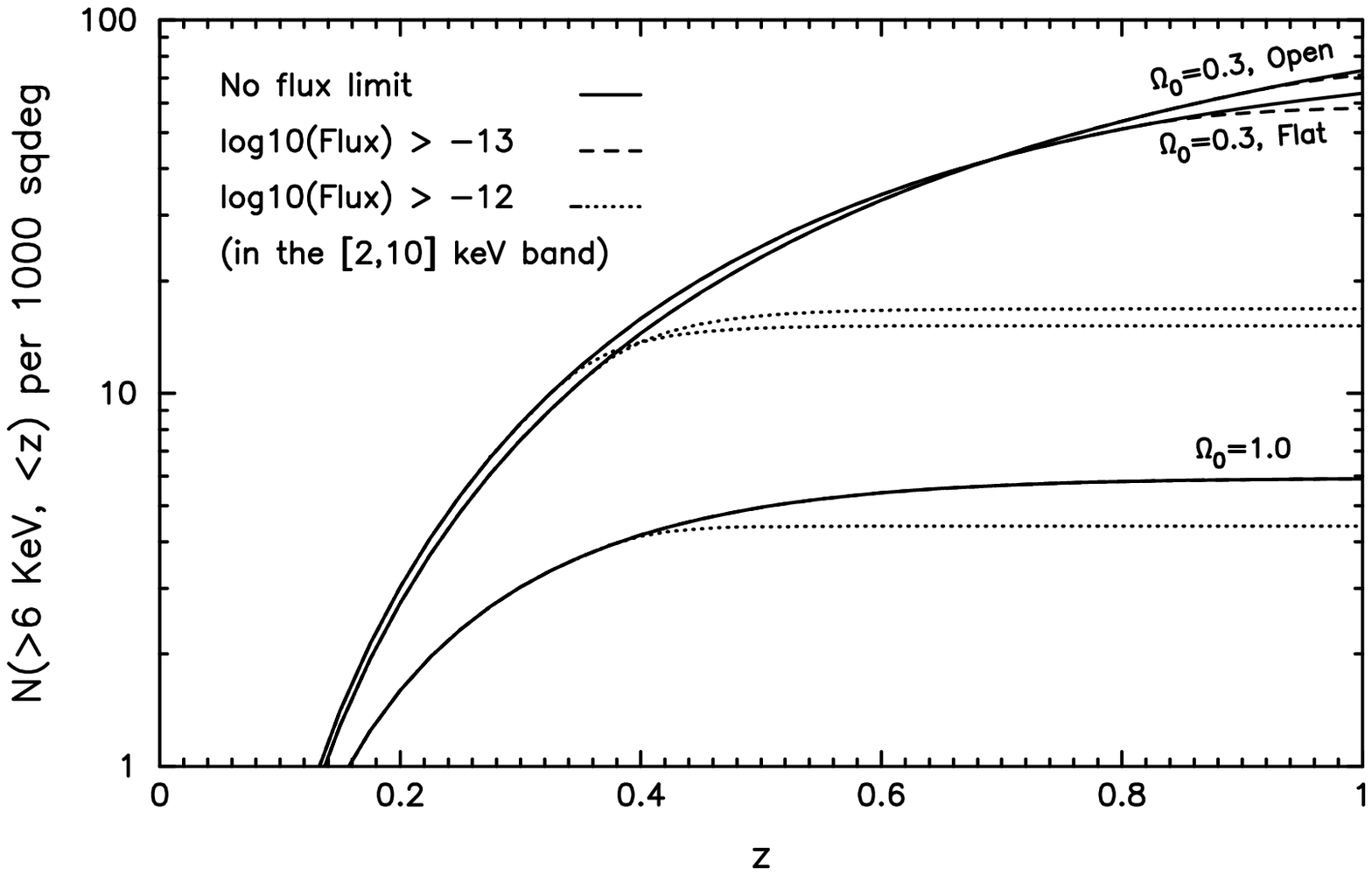}}
\centerline{\includegraphics[width=3.25in]{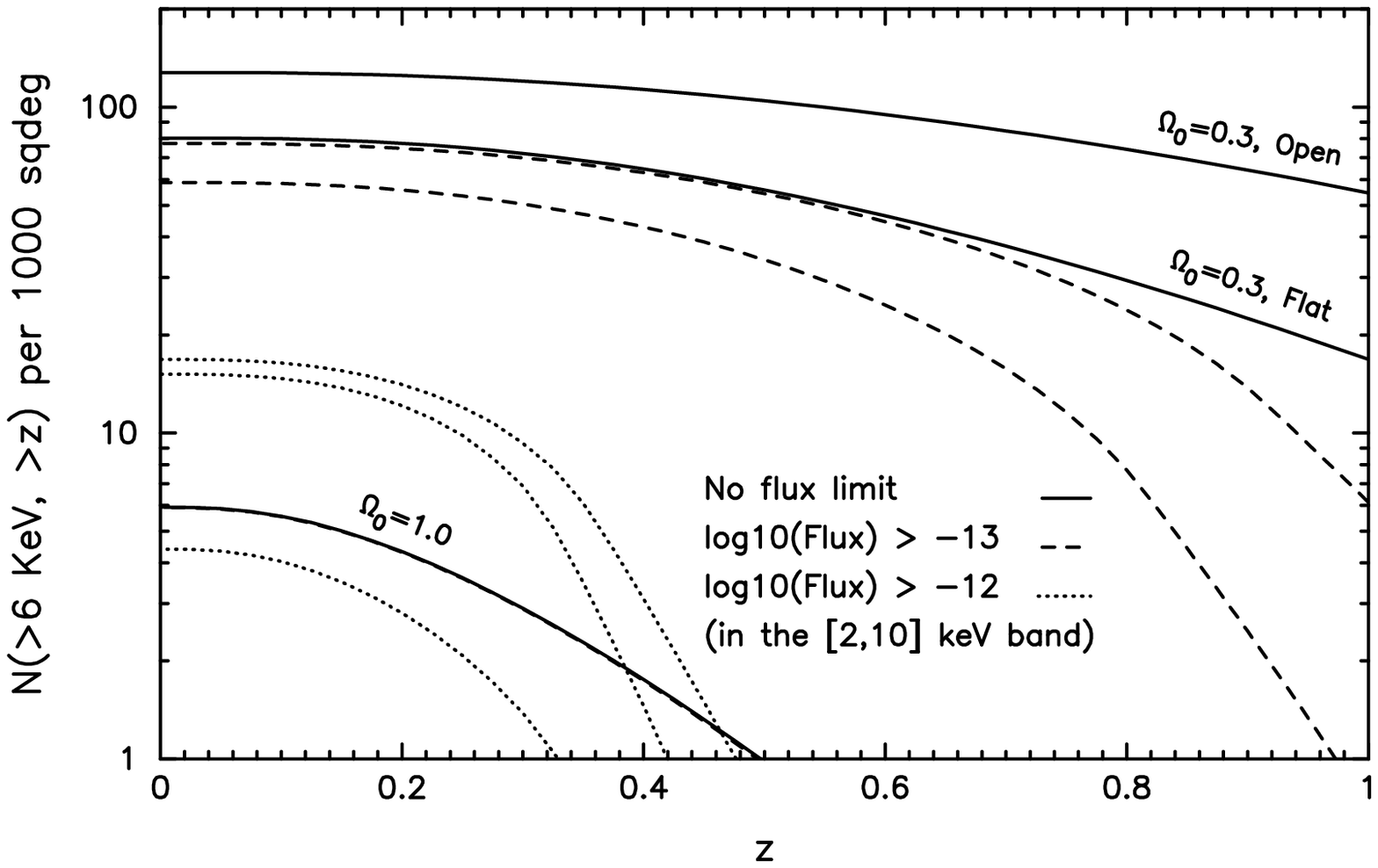}}
\figcaption{The cumulative redshift distribution, out to redshift $z$ 
(upper panel) and beyond redshift $z$ (lower panel), of 
the number of galaxy clusters per 1000 square degrees with X-ray 
temperature in excess of 6 keV. The solid lines show the result 
for all clusters, while the dashed and dotted lines represent only the 
clusters which have a X-ray flux in the [2,10] keV band in excess of 
$10^{-13}$ and $10^{-12}\;{\rm erg}\,{\rm cm}^{-2}\,{\rm s}^{-1}$, 
respectively. Each curve is normalized to reproduce the observed present 
abundance of high-temperature galaxy clusters.}  
\medskip

It is clear that the predictions of models with different $\Omega_{0}$ start to 
differ very rapidly as the redshift probed is increased. When $z$ reaches 
0.3, in principle it would be already possible to distinguish between models 
with substantially different values for $\Omega_{0}$. As one considers data 
pertaining to increasingly higher redshifts, the error bar on the estimation 
of $\Omega_{0}$ becomes ever smaller. At redshifts above 1, it starts to be 
even possible to determine whether in the case of a low $\Omega_{0}$ universe, 
there is a cosmological constant making it spatially-flat. 

Unfortunately, there are two limitations. The first is instrumental noise, as all 
detectors have a flux threshold below which the signal-to-noise becomes 
sufficiently small to cast serious doubts over any detection. Further, given that 
we are interested in surveys where the clusters have their X-ray temperatures 
measured, in practice the X-ray flux threshold of a cluster 
catalogue with temperatures 
will be around an order of magnitude larger than the X-ray flux detection 
threshold of the survey from which the catalogue was assembled. This problem 
means that surveys with high X-ray flux thresholds will not include many 
clusters at high redshifts with their X-ray temperatures measured. 
In Figure 4, one can see that a cluster catalogue with measured X-ray 
temperatures assembled by imposing a X-ray flux threshold of 
$10^{-12}\;{\rm erg}\,{\rm cm}^{-2}\,{\rm s}^{-1}$ will most 
probably not include any cluster with $z>0.4$, even if the area of the 
survey covered a large portion of the sky. A survey an order of magnitude more 
sensitive in flux would include almost all clusters with 
$k_{{\rm B}}T>6\,{\rm keV}$ up to a redshift of 1, and several beyond if they 
exist. 

The second problem is the uncertainties in the theoretical modelling of the 
structure and number density evolution of galaxy clusters, and also the still 
significant error on the present-day cluster abundance. The last issue may 
become less of a problem once data for several redshift intervals becomes 
available. These uncertainties were sufficient to make the presently available 
cluster abundance data for $0.3<z<0.4$ not very useful in the determination 
of $\Omega_{0}$. Given that a XMM Slew Survey 
(Lumb 1998; also see Laurence Jones in these proceedings),  
because of its short exposures, would only be able to measure X-ray temperatures  
for clusters with fluxes exceeding about 
$10^{-12}\;{\rm erg}\,{\rm cm}^{-2}\,{\rm s}^{-1}$, 
almost all such clusters would have $z<0.4$, and thus such a survey 
would probably turn out not very useful in the estimation of $\Omega_{0}$. 
However, a XMM Serendipitous Survey, though it would cover only about one tenth 
of the sky area that could be covered by a XMM Slew Survey, it would be an 
order of magnitude more sensitive. All clusters with X-ray fluxes in excess of 
about $10^{-13}\;{\rm erg}\,{\rm cm}^{-2}\,{\rm s}^{-1}$, 
which as we mentioned above in practice 
means almost all clusters with X-ray temperatures in excess of 6 keV up to a 
redshift of 1, would have their X-ray temperatures measured (Romer 1998). 
A catalogue built around these clusters would already go deep enough in redshift 
to allow a good estimate of $\Omega_{0}$, even taking into account the theoretical 
modelling uncertainties. 

\acknowledgements{We are very grateful to Patrick Henry, Alain 
Blanchard, Vincent Eke and Monique Arnaud for many useful 
comments and discussions. PTPV is supported 
by the PRAXIS XXI program of FCT (Portugal), and ARL by the Royal Society.}

\end{document}